\pdfoutput=1
\documentclass[cernpreprint, atlasdraft=false, UKenglish, texmf, orcidlogo]{atlasdoc}
\usepackage{atlaspackage}
\usepackage{atlasbiblatex}

\usepackage{atlasphysics}
\graphicspath{{logos/}{figures/}}

\usepackage{multirow}

\AtlasTitle{Search for the associated production of charm quarks and a Higgs boson decaying into a photon pair with the ATLAS detector}

\AtlasAbstract{A search for the production of a Higgs boson and one or more charm quarks, in which the Higgs boson decays into a photon pair, is presented. This search uses proton--proton collision data with a centre-of-mass energy of $\sqrt{s}=13$~TeV and an integrated luminosity of 140~fb$^{-1}$ recorded by the ATLAS detector at the Large Hadron Collider. The analysis relies on the identification of charm-quark-containing jets, and adopts an approach based on Gaussian process regression to model the non-resonant di-photon background. The observed (expected, assuming the Standard Model signal) upper limit at the 95\% confidence level on the cross-section for producing a Higgs boson and at least one charm-quark-containing jet that passes a fiducial selection is found to be 10.6~pb (8.8~pb). The observed (expected) measured cross-section for this process is $5.3 \pm 3.2~\text{pb}$ ($2.9 \pm 3.1~\text{pb}$).}

\AtlasRefCode{HIGG-2021-06}

\PreprintIdNumber{CERN-EP-2024-175}

\AtlasJournal{JHEP}
\AtlasJournalRef{JHEP 02 (2025) 045}
\AtlasDOI{10.1007/JHEP02(2025)045}

\hypersetup{pdftitle={ATLAS document},pdfauthor={The ATLAS Collaboration}}

\begin{document}

\maketitle

\section{Introduction and signal definition}\label{sec:intro}

The Higgs boson is the only scalar elementary particle in the Standard Model (SM) of particle physics, and is associated with a field with a non-zero vacuum expectation value. It arises from the Brout-Englert-Higgs mechanism~\cite{Higgs:1964ia,Higgs:1964pj,Guralnik:1964eu,PhysRev.145.1156,Kibble:1967sv,Englert:1964et}, which generates the masses of all known fundamental particles except possibly for the neutrinos.
Since its discovery in 2012~\cite{HIGG-2012-27,CMS-HIG-12-028}, an extensive campaign of measurements~\cite{HIGG-2021-23,CMS-HIG-22-001} has been underway at the Large Hadron Collider (LHC)~\cite{Evans:2008zzb} to understand its properties and test the SM.
The Yukawa sector of the SM~\cite{Weinberg:1967tq} does not explain the structure of the fermion masses, and only predicts that their couplings to the Higgs boson are proportional to their masses. These masses take on hierarchical values, and span a wide range. This hierarchy may be explained by a mechanism beyond the SM~\cite{FROGGATT1979277,GIUDICE200879,deFlorian:2016spz}.
Measuring these couplings is thus a crucial test of the SM, and can be sensitive to new physics.
In particular, significant work has been dedicated to probing the coupling of the Higgs boson to charm quarks ($y_c$)~\cite{HIGG-2021-12,CMS-HIG-21-008,HDBS-2018-53,HIGG-2022-04,CMS-HIG-17-028,CMS-SMP-17-012}, though this has so far been difficult due to its relatively low value in the SM and significant hadronic backgrounds at the LHC.

This paper presents the first search for inclusive Higgs boson plus charm-quark production (the inclusive $H+c$ process), which uses decays of the Higgs boson to photon pairs, and the ATLAS detector at the LHC\@.
A component of the inclusive $H+c$ process is a $g+c\to H+c$ process that is sensitive to $y_c$ (the $y_c$-sensitive $H+c$ process).
Despite the $y_c$-sensitive $H+c$ process amounting only to approximately 1\% of the inclusive $H+c$ process, the quadratic dependence of its cross-section on $y_c$ could lead to important effects on these signals in the case of deviations from the SM\@. This measurement thus provides an important step towards probing $y_c$ using $H+c$ events~\cite{PhysRevLett.115.211801}, which complements the methods cited above.

The inclusive $H+c$ signal targeted herein is defined as any event including a Higgs boson and at least one anti-$k_{t}$ jet~\cite{Cacciari:2008gp,Fastjet} with a radius parameter $R=0.4$ formed from generator-level particles, which has $p_{\mathrm{T}}>25$~\GeV and $|\eta|<2.5$ and is matched to a charm hadron with $p_{\mathrm{T}}>5$~\GeV and $\Delta R <0.3$, excluding the cascade decays from $b$-hadrons. Various Higgs boson production processes contribute to this signal, and the simulation of these processes is described in Section~\ref{sec:detdatsim}. The backgrounds considered are the dominant non-resonant $pp\rightarrow \gamma\gamma + n$ parton production background, and the largely irreducible Higgs boson-induced resonant background. The non-resonant background is estimated by using a data-driven approach. Events containing a Higgs boson that do not satisfy the signal criteria are considered as resonant backgrounds, these often have light-flavour-containing jets (light-flavour jets) or bottom-quark-containing jets ($b$-jets) mis-identified as charm-quark-induced jets ($c$-jets), and they are modelled using simulation. %

\section{The ATLAS detector, data and simulation}
\label{sec:detdatsim}

Proton-proton collision data collected at a centre-of-mass energy of $\sqrt{s}=13$~\TeV during Run~2 of the LHC, with an integrated luminosity of 140~fb$^{-1}$, are used. The data are recorded by the ATLAS detector~\cite{PERF-2007-01}, a multipurpose detector with forward-backward symmetric cylindrical geometry and nearly $4\pi$ coverage in solid angle\footnote{ATLAS uses a right-handed coordinate system with its origin at the nominal interaction point (IP) in the centre of the detector and the \(z\)-axis along the beam pipe. The \(x\)-axis points from the IP to the centre of the LHC ring, and the \(y\)-axis points upwards. Polar coordinates \((r,\phi)\) are used in the transverse plane, \(\phi\) being the azimuthal angle around the \(z\)-axis. The pseudorapidity is defined in terms of the polar angle \(\theta\) as \(\eta = -\ln \tan(\theta/2)\) and is equal to the rapidity $ y = \frac{1}{2} \ln \left( \frac{E + p_z c}{E - p_z c} \right) $ in the relativistic limit. Angular distance is measured in units of \(\Delta R \equiv \sqrt{(\Delta y)^{2} + (\Delta\phi)^{2}}\).}. ATLAS consists of an inner tracking detector, electromagnetic and hadronic calorimeters, and a muon spectrometer equipped with a two-level trigger system~\cite{TRIG-2016-01}. An extensive software suite~\cite{SOFT-2022-02} is used in data simulation, in the reconstruction and analysis of real and simulated data, in detector operations, and in the trigger and data acquisition systems of the experiment. The data were collected using di-photon and single-photon triggers, which have various minimum transverse energy ($E_{\mathrm{T}}$) thresholds. Di-photon triggers are used, which require events to contain at least two photons that satisfy identification and isolation requirements that vary by data-taking year, and which have minimum $E_{\mathrm{T}}$ thresholds of 35~(25)~\GeV for the (second) highest $E_{\mathrm{T}}$ photon. Single-photon triggers with a photon identification requirement and with $E_{\mathrm{T}}$ thresholds of 120~\GeV (140~\GeV) are used to select the data collected in 2015 (2016--2018)\cite{TRIG-2018-05}.

\begin{figure*}[htbp]
\begin{center}
\subfloat[]{\includegraphics[width=0.32\textwidth]{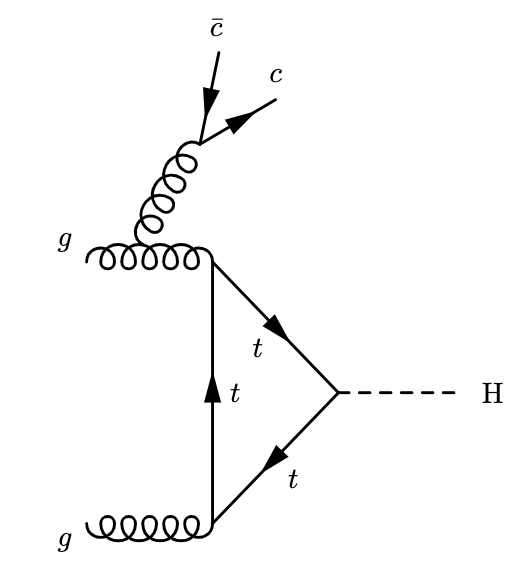}}\hspace{0.4cm}
\subfloat[]{\includegraphics[width=0.3\textwidth]{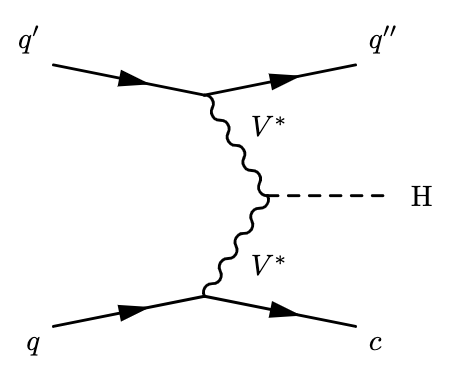}}\\
\subfloat[]{\includegraphics[width=0.3\textwidth]{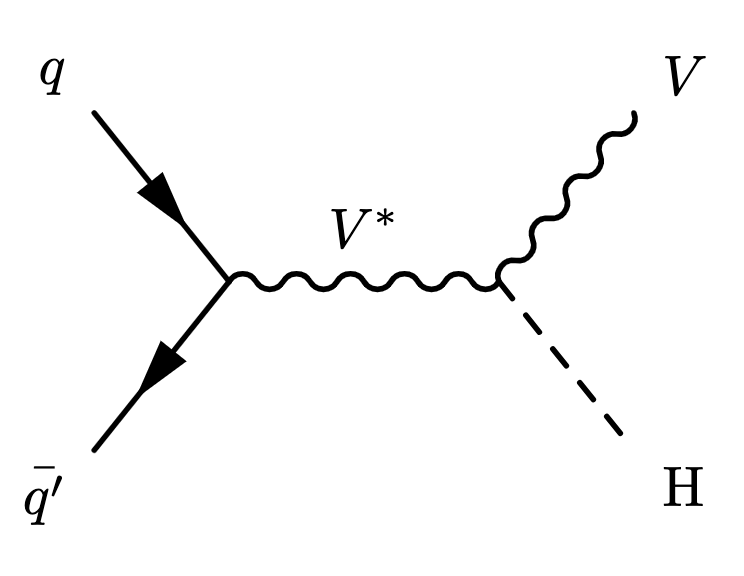}}\hspace{1.5cm}
\subfloat[]{\includegraphics[width=0.3\textwidth]{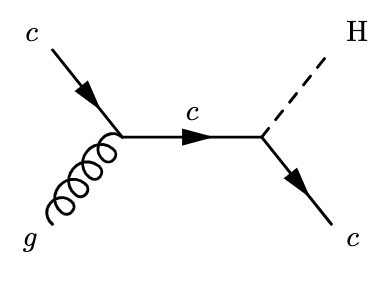}}
\caption{Example Feynman diagrams for processes contributing to the inclusive $H+c$ signal: (a) gluon-gluon fusion Higgs boson production with gluon radiation, which splits into $c\bar{c}$; (b) vector boson fusion Higgs boson production; (c) the $VH$ processes, where $V$ can be a $W$ or $Z$ boson decaying into a final state with one charm quark or a pair, respectively; and, (d) the $s$-channel diagram for the $y_c$-sensitive $H+c$ signal.} %
\label{fig:dia}
\end{center}
\end{figure*}

The inclusive $H+c$ signal and the resonant backgrounds containing a Higgs boson are simulated together, and then separated using a generator-level implementation of the signal definition stated above. The Higgs boson production modes simulated for these combined samples are gluon-gluon fusion (ggF), vector boson fusion (VBF), associated production with a vector boson ($ZH$ and $WH$), with a top-quark pair ($t\bar{t}H$) and with a bottom-quark pair ($b\bar{b}H$), and the $y_c$-sensitive $H+c$ process. Some example Feynman diagrams of these processes are shown in Figure~\ref{fig:dia}. The inclusive $H+c$ signal cross-section is about 2.9~pb in the SM, as determined using the simulation method described below, though this value is uncertain due to the difficulty in calculating the cross-section for Higgs bosons produced with jets containing a $c$-quark or $b$-quark (heavy-flavour jets). The event generation is performed using \POWHEGBOX~v2~\cite{Frixione:2007vw,Alioli:2010xd,Alioli:2008tz,Nason:2004rx} for all production modes, other than the $y_c$-sensitive $H+c$ process, which utilises \MADGRAPH5~\cite{Alwall_2011}, using matrix element calculations at the highest available order of accuracy in the strong coupling constant $\alpha_s$. The ggF sample is generated at the next-to-next-to-leading order (NNLO), while next-to-leading order (NLO) accuracy is achieved for VBF, $WH$, $q\bar{q}\rightarrow ZH$, $t\bar{t}H$ and $b\bar{b}H$. The $gg\rightarrow ZH$ simulation is performed at leading order. The $y_c$-sensitive $H+c$ sample models the $g+c\rightarrow H+c$ process in which the Higgs boson couples directly to a charm quark, and it is simulated using \MADGRAPH5 at LO and Higgs Effective Couplings UFO~\cite{DEGRANDE20121201}, where generator-level jets are required to have $p_{\mathrm{T}}>10$~\GeV and $|\eta|< 4.7$. All samples are interfaced with \Pythia8 \cite{Sjostrand:2007gs,Sjostrand:2014zea} to simulate the hadronisation, parton shower and underlying event. The PDF4LHC15 parton distribution function (PDF) set~\cite{Butterworth:2015oua} and AZNLO CTEQ6L1 tuning parameters~\cite{STDM-2012-23,Ball:2014uwa} are used for all samples except for $t\bar{t}H$ and $b\bar{b}H$, which use the A14 \NNPDF[2.3lo] tune~\cite{ATL-PHYS-PUB-2014-021,Ball:2014uwa}, and for the $y_c$-sensitive $H+c$ sample, which uses \NNPDF[3.0nlo]~\cite{Ball:2014uwa} and the A14 \NNPDF[2.3lo] tune. The $H\to\gamma\gamma$ decays are modelled using \Pythia8 and include the small contribution from Dalitz decays ($H\rightarrow \gamma f \bar{f}$). In all samples, the Higgs boson mass and width are set to 125~\GeV and 4.07~\MeV, respectively, and the normalisations (except for the $y_c$-sensitive $H+c$ sample) use the latest theoretical calculations for SM production cross-sections~\cite{deFlorian:2016spz}. The $y_c$-sensitive $H+c$ sample is normalised to the SM production cross-section of 0.028~pb, which is obtained from \MADGRAPH5 using $m_c(m_Z)=0.63$~\GeV. The branching fraction of Higgs boson decay into a photon pair ($H\rightarrow\gamma\gamma$) is assumed to be $2.27^{+0.07}_{-0.06} \times 10^{-3}$~\cite{deFlorian:2016spz}. The response of the ATLAS detector is modelled using \GEANT~\cite{Agostinelli:2002hh,SOFT-2010-01}. The dominant non-resonant $pp\rightarrow \gamma\gamma + n$ parton production background is simulated to validate and evaluate an uncertainty in the data-driven method used to estimate this background. This sample is produced using \Sherpa~2.2.4~\cite{Sherpa:2019gpd} with NLO-accurate matrix elements for up to one parton, and LO accurate matrix elements for up to three partons, and was processed using a parameterised simulation of the ATLAS detector. All simulated events are reconstructed using the same algorithms that are applied to data~\cite{SOFT-2010-01}, and include the effect of multiple proton-proton interactions per bunch crossing (pile-up), as well as the effect on the detector response due to interactions from bunch crossings before or after the one containing the hard interaction.

\section{Event reconstruction and selection}

The event reconstruction and selection closely follows the recent Higgs boson production cross-section measurement analysis in the di-photon decay channel~\cite{HIGG-2020-16}. Photons are reconstructed from energy deposits in the calorimeter formed using a topological cell clustering algorithm~\cite{EGAM-2018-01}. The event selection is performed in two stages. The first stage constitutes a preselection, where the two highest-$E_\text{T}$ preselected photon candidates are required to satisfy $E_\text{T} > 25~\text{\GeV}$,  $|\eta|<2.37$, excluding the transition region between the barrel and endcap electromagnetic calorimeters $1.37<|\eta|<1.52$, and to satisfy $loose$ calorimeter-based identification criteria~\cite{EGAM-2021-01_mod}. The accurate determination of the di-photon production vertex is crucial for measuring the di-photon invariant mass $m_{\gamma\gamma}$, selecting jets from the hard interaction, and calculating track-based isolation. To facilitate this, the preselected photon angular information and reconstructed vertex information are input to a neural network that is trained on simulated events to improve the determination of the primary vertex~\cite{HIGG-2013-08}. This vertex is then used to calculate properties of physics objects in the event. In the second selection stage, the two preselected photon candidates are required to satisfy $tight$ identification criteria~\cite{EGAM-2021-01_mod}. In addition, candidates must meet calorimeter- and track-based isolation criteria to minimise the misidentification of jets as photons. Calorimeter-based isolation refers to the total transverse energy of calorimeter clusters within a cone of size $\Delta R =0.2$ around the photon candidate. This excludes the energy in a fixed-size window containing the photon shower; the corrections for leakage, as well as subtraction of pile-up and underlying event contribution, are also performed~\cite{EGAM-2018-01}. The calorimeter-based isolation must be under 6.5\% of the transverse energy of the photon. The track-based isolation is defined as the scalar sum of the transverse momenta of tracks within a cone of size $\Delta R =0.2$ around the photon candidate.
Only the tracks with transverse momenta $p_\text{T}>1~\text{\GeV}$ linked to the di-photon vertex and excluding those associated with photon conversions are used. The track isolation must be less than 5\% of the photon transverse energy. The invariant mass of the di-photon system $m_{\gamma\gamma}$ must lie between 105 and 160~\GeV. Finally, the leading (subleading) photon is required to satisfy $E_\text{T}/m_{\gamma\gamma} > 0.35~(0.25)$~\cite{HIGG-2013-12}.

Jet constituents are reconstructed using a particle-flow algorithm~\cite{PERF-2015-09}, from which jets are reconstructed using the anti-$k_t$ algorithm with a radius parameter of $R = 0.4$. Events are required to have one or two jets with $p_\text{T}>25~\text{\GeV}$ and $|\eta|<2.5$, and each jet is required to have $\Delta R(j,\gamma)>1$ from both of the photons. The jets with $|\eta|<2.4$ and $p_\text{T} <$~60~\GeV, originating from additional $pp$ collisions in the same or neighbouring bunch crossings are suppressed by the use of the jet-vertex-tagger~\cite{PERF-2014-03}.  The flavour of the quark that initiates each reconstructed jet is identified using a deep learning-based algorithm called DL1r~\cite{FTAG-2019-07}, which estimates the probabilities of a jet being a $b$-jet, a $c$-jet  or a light-flavour jet. The output probabilities are combined with a parameter that regulates the discrimination between $c$-jets and $b$-jets or light-flavour jets. Finally, a requirement is made on this parameter to determine whether a jet is considered as $c$-tagged. The average efficiencies are 30--40\%, 10--13\% and 1--2\% for jets originating from $c$-quarks, $b$-quarks and light-flavour quarks, respectively, depending on the Higgs boson production mode~\cite{TOPQ-2019-04}. The overlap removal procedure described in Ref.~\cite{HIGG-2020-16} is applied to avoid double-counting objects. To optimise the analysis sensitivity, selected events are split into two signal regions (SRs): the $c$-tag SR, where at least one $c$-tagged jet is present; and the non-$c$-tag SR, where no $c$-tagged jet is present. The non-$c$-tag SR provides significant sensitivity due to the low efficiency of the $c$-tagging criterion. Lastly, for an event to fall in a SR it must satisfy $120~\text{\GeV}<m_{\gamma\gamma}<130~\text{\GeV}$, while events not meeting this requirement fall into the sidebands, which are later used to estimate the non-resonant background. The approximate expected yields of different physics processes contributing to the signal and resonant background are shown in Table~\ref{tab:yields}, and these numbers are meant for illustrative purposes.

\begin{table}[!htbp]
\centering
\caption{Approximate expected yields of different physics processes contributing to the Standard Model signal and Standard Model Higgs boson-induced resonant background, in the $c$-tag and non-$c$-tag signal regions, for an integrated luminosity of 140 fb$^{-1}$. The $y_c$-sensitive $H+c$ process assumes the Standard Model value of the coupling between the Higgs boson and charm quarks. The signal and background yields for the individual processes are estimated by using the simulation methods detailed in the main text, and the Total corresponds to the sum of the signal and resonant background processes.} %
\begin{tabular}{l S[table-format=2.0] c S[table-format=2.0] c S[table-format=2.0] c S[table-format=2.0] c}
\toprule
\multirow{2}{*}{Process}               & \multicolumn{3}{c}{$c$-tag signal region} &  & \multicolumn{3}{c}{Non-$c$-tag signal region} & \\
& {Signal}                      &  & {Resonant background}  &  & {Signal} &  & {Resonant background} & \\
\midrule
ggF~$H$               & 39                                    &  & 82                &  & 110                  &  & 1800                & \\ %
VBF~$H$               & 17                                    &  & 13                &  & 34                   &  & 220                 & \\ %
$WH$                  & 9.5                                   &  & 4.7               &  & 23                   &  & 59                  & \\
$ZH$                  & 4.5                                   &  & 5.1               &  & 7.8                  &  & 50                  & \\
$t\bar{t}H$           & 7.0                                   &  & 4.6               &  & 20                   &  & 24                  & \\
$b\bar{b}H$           & 0.11                                  &  & 1.9               &  & 0.35                 &  & 16                  & \\
$y_c$-sensitive $H+c$ & 0.37                                  &  & 0.046             &  & 0.78                 &  & 0.48                & \\
Total                 & 77									 &  & 110               &  & 190                  &  & 2100                & \\
\bottomrule
\end{tabular}\\
\label{tab:yields}
\end{table}

\section{Signal and background modelling}

\begin{figure}[htbp]
\begin{center}
\subfloat[]{\includegraphics[width=0.5\textwidth]{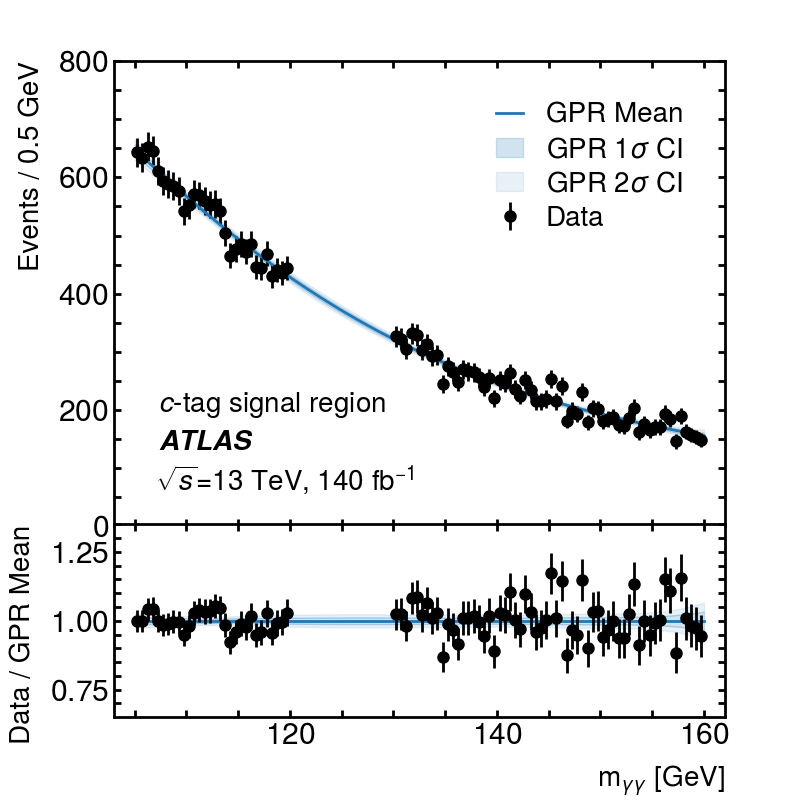}}
\subfloat[]{\includegraphics[width=0.5\textwidth]{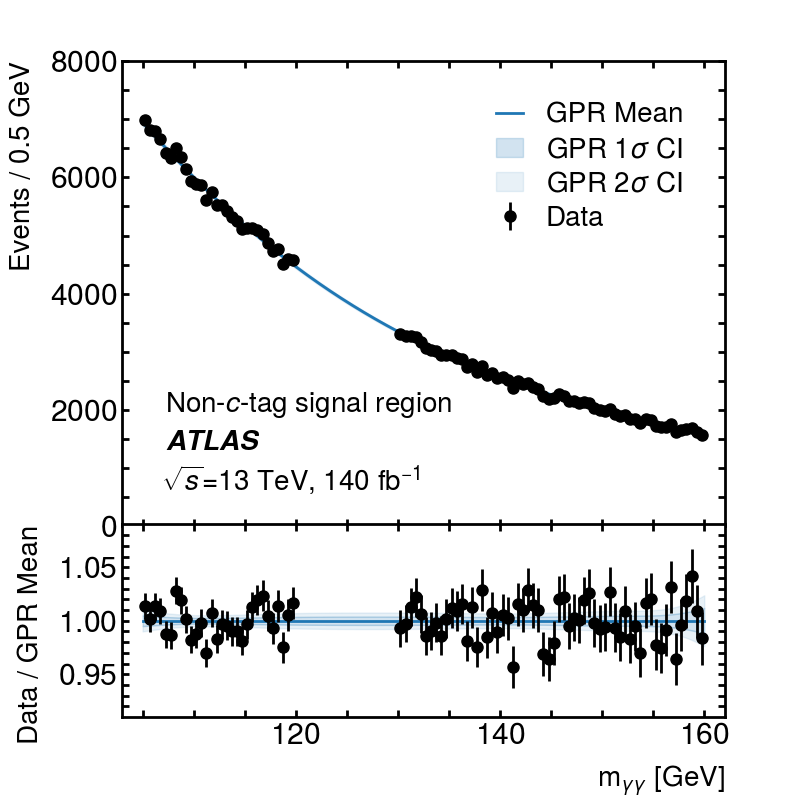}}
\caption{Gaussian process regression-based estimates of the non-resonant background derived from the data in the $m_{\gamma\gamma}$ sidebands as detailed in the main text in the (a) $c$-tag and (b) non-$c$-tag combined signal regions and sidebands. The $1\sigma$ and $2\sigma$ Bayesian credible intervals (CI) are shown by the darker and lighter shaded regions, respectively. The data in the sidebands, but not in the signal regions, is also shown.}
\label{fig:gpr}
\end{center}
\end{figure}

\begin{figure}[htbp]
\begin{center}
\subfloat[]{\includegraphics[width=0.5\textwidth]{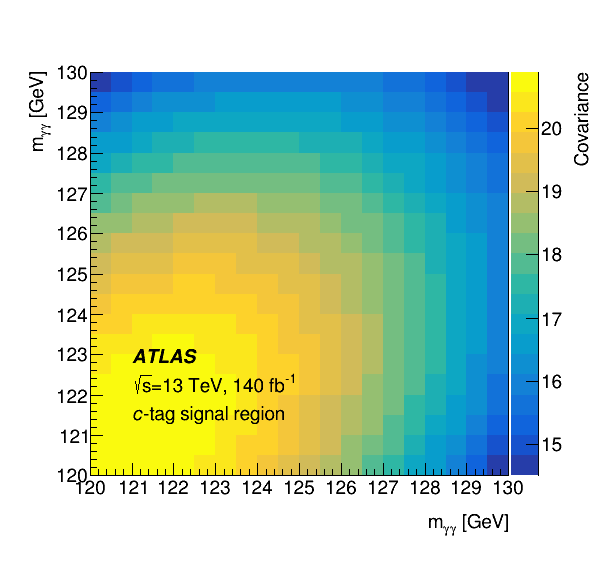}}
\subfloat[]{\includegraphics[width=0.5\textwidth]{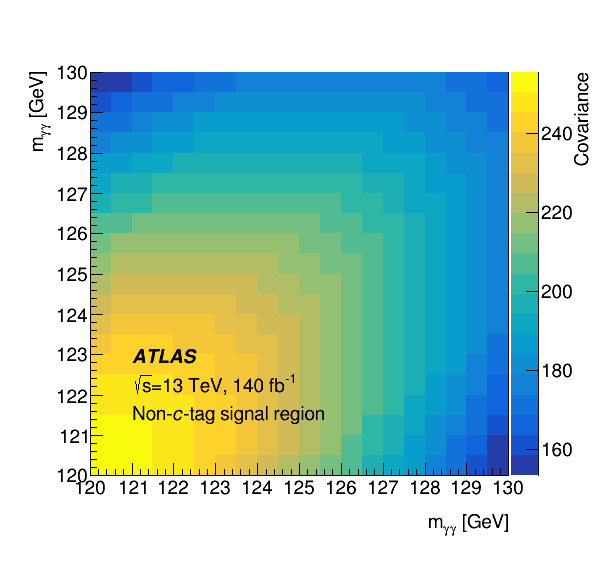}}
\caption{Covariance matrices of the Gaussian process regression-based estimates of the non-resonant background derived from the data in the $m_{\gamma\gamma}$ sidebands as detailed in the main text in the (a) $c$-tag and (b) non-$c$-tag signal regions.}
\label{fig:covs}
\end{center}
\end{figure}

The inclusive $H+c$ signal and the total resonant and non-resonant background models are constructed in both of the SRs as $m_{\gamma\gamma}$ distributions with 0.5~\GeV bins, in the mass range between 120~\GeV and 130~\GeV. The signal and total resonant background are modelled in each SR using simulation, as described above. The non-resonant background estimates employ a data-driven approach using Gaussian process regression (GPR)~\cite{Krige1951ASA} to interpolate from the data in the sidebands, defined as lower ($105<m_{\gamma\gamma}<120$~\GeV) and upper ($130<m_{\gamma\gamma}<160$~\GeV) mass regions, into the SRs. GPR is a non-parametric method that provides a Bayesian posterior distribution in the form of a functional over possible background functions. This approach provides a highly flexible background estimate, which explores a complete space of possible background shapes, and which has limited dependence on the choice of the prior distribution over possible background distributions~\cite{ATL-PHYS-PUB-2020-028}. Other implementation methods have been proposed in the particle physics literature~\cite{Frate:2017mai} that use sidebands and the signal regions simultaneously to derive the GPR estimate, however, here a blinding and interpolation-based approach using the sidebands only is adopted. Histograms with 0.5~\GeV bins in the sidebands are used to produce estimates of the non-resonant background distributions in their respective SRs using GPR-based interpolation. The GPR estimates use a Radial Basis Function kernel~\cite{10.7551/mitpress/3206.001.0001}, which is motivated by the smoothness of these backgrounds, as is evidenced by both the simulation and the data sidebands. The histograms input to the GPR algorithm are pre-processed, and the parameters of the kernels are optimised by maximising the log-marginal-likelihoods of the data in the respective sidebands. GPR is then implemented using \texttt{scikit-learn}~\cite{scikit-learn} to provide posterior distributions over possible background distributions in the SRs, which are represented by multidimensional Gaussian distributions over the histogram bin contents. The means of the $m_{\gamma\gamma}$ histogram bin contents arising from the GPR estimates are shown in Figure~\ref{fig:gpr}, and the covariances of the bin contents in the SRs are shown in Figure~\ref{fig:covs}. These posterior distributions are used to model the non-resonant background and the uncertainties arising from the GPR process in the statistical interpretation. These uncertainties arise from the GPR prior, the limited amount of data in the sidebands and the interpolations from the sidebands into the SRs, and they are largely statistical in nature. This method assumes that there is negligible signal in the sidebands, and this is confirmed to be the case using simulation. The robustness of the GPR-based background modelling strategy is confirmed using the \Sherpa $\gamma\gamma + n$ parton simulated sample described above. %

\section{Systematic uncertainties}

The inclusive $H+c$ signal and total resonant background, which are both modelled using simulation, are impacted by various theoretical and experimental systematic uncertainties. The data-driven non-resonant background model is unaffected by these uncertainties, but affected by dedicated modelling uncertainties described below.

Theoretical uncertainties in the signal and resonant background are assessed by varying the renormalisation and factorisation scales, and by calculating the impacts of uncertainties in the PDF, $\alpha_s$ and the branching fraction. Theoretical uncertainties related to the parton shower, hadronisation and underlying events are evaluated by comparing with results obtained from the alternative generator \texttt{Herwig}~7.1.3~\cite{Bellm:2017bvx}. A 100\% normalisation uncertainty is applied to resonant background events containing Higgs bosons with heavy-flavour jets from the ggF or VBF processes, to account for the difficulty in modelling these processes for events containing heavy-flavour jets, since these jets do not typically arise from the hard-scatter. This is motivated by data-to-simulation differences observed in other analyses, e.g., Refs.~\cite{HIGG-2019-13,HIGG-2018-29}. The signal efficiency has an overall theoretical uncertainty of less than 12\% in both the SRs, while the theoretical uncertainty in the resonant background is less than 18\% and 8\% in the $c$-tag and non-$c$-tag SRs, respectively.

Experimental uncertainties are applied to the photon identification and isolation efficiencies and their energy scale and resolution~\cite{EGAM-2018-01}, of the jet vertex tagger efficiency~\cite{PERF-2014-03}, energy scale and resolution~\cite{JETM-2018-05}, and of the $c$-tagging efficiency~\cite{TOPQ-2019-04} and fake rates. An uncertainty is assigned to account for the different primary vertex definitions used in this analysis and the $c$-tagging calibration, which results in uncertainties of (16\%) 16\% for signal and (3.9\%) 16\% for resonant background in the (non-)$c$-tag SR\@. These uncertainties are uncorrelated between the signal and resonant backgrounds, and anti-correlated between the SRs. An additional uncertainty is assigned to events for which the $c$-tagged jet arises from the mis-tagging of a light-flavour jet to account for the fact that the calibration of the corresponding mis-tag rate is performed using $t\bar t$ events, which do not contain a highly pure component of light-flavour jet events, unlike the samples that are usually adopted for this calibration task~\cite{FTAG-2019-02}. This uncertainty is designed to be approximately twice as large as it would need to be to make the scale factor from the calibration compatible with unity for light-flavour jets, providing a conservative uncertainty, and resulting in uncertainties of (0.36\%) 0.72\% for the signal and (2.2\%) 30\% for the resonant background in the (non-)$c$-tag SR\@. Moreover, uncertainties in luminosity~\cite{DAPR-2021-01,LUCID2} and trigger efficiency~\cite{TRIG-2016-01} are considered, which would impact the event yield by 0.83\% and 0.4\%, respectively. An uncertainty in the modelling of pile-up events is included, which is less than 1.1\%~\cite{STDM-2015-05}. The main part of the uncertainty in the non-resonant background is estimated using the GPR method. To account for any uncertainty not accounted for by the GPR method itself (e.g. from the choice of GPR prior), a spurious signal test is performed using an approach similar to that described in Ref.~\cite{ATL-PHYS-PUB-2020-028}. This spurious signal test uses pseudo-data events generated based on the non-resonant simulated \Sherpa sample with data-driven corrections~\cite{HIGG-2016-21} that account for jets being mis-identified as photons, and results in an uncertainty of 12\% (30\%)  of the expected signal in the $c$-tag (non-$c$-tag) category. This spurious signal uncertainty is an additive pre-fit normalisation uncertainty, while the other normalisation-only uncertainties mentioned in this paragraph are multiplicative pre-fit normalisation uncertainties. The photon energy scale and resolution uncertainties and the GPR-derived background estimation uncertainties affect the normalisations and shapes of the $m_{\gamma\gamma}$ distributions, while the other systematic uncertainties affect only the normalisations.

Uncertainties that do not affect the normalisation of the signal or the total resonant background by more than 1\% in either SR are removed from the analysis. The breakdown of the expected inclusive $H+c$ signal uncertainty into various categories of uncertainty in this search, as obtained after the fit to the Asimov dataset described in Section~\ref{sec:stats}, is shwon in Table~\ref{tab:breakdown}.

\begin{table}[!htbp]
\centering
\caption{Breakdown of the expected relative signal uncertainty into various categories, calculated by subtracting the uncertainty in the signal cross-section (with the relevant nuisance parameters fixed) in quadrature from the total signal uncertainty (with these nuisance parameters allowed to vary), and then dividing by the total uncertainty. The GPR posterior uncertainty is largely, but not entirely, statistical in nature. The ``Theory'' category consists of parton shower algorithm, parton distribution function, renormalisation and factorisation scale, the value of the strong coupling constant ($\alpha_s$), $H\rightarrow \gamma\gamma$ branching fraction and Higgs boson$ + \text{heavy-flavour jets}$ cross-section uncertainties.}
\begin{tabular}{lc}
\toprule
Uncertainty & $H+c$ uncertainty impact \\
\midrule
\textbf{Statistical (incl. GPR)} & 78\% \\
~~~GPR posterior & 46\% \\
\textbf{Systematic (excl. GPR)} & 63\% \\
~~~Theory & 43\% \\
~~~$c$-tagging & 29\% \\
~~~Photons & 27\% \\
~~~Jets & 20\% \\
~~~Spurious signal & 11\% \\
~~~Pile-up & 2\% \\
\bottomrule
\end{tabular}\\
\label{tab:breakdown}
\end{table}

\section{Statistical interpretation}
\label{sec:stats}

The statistical interpretation uses a binned likelihood fit to the $m_{\gamma\gamma}$ distributions, simultaneously in both the SRs in the range of 120--130~\GeV. The signal and total resonant background are each modelled using one histogram per SR\@. The non-resonant background estimates consist of $m_{\gamma\gamma}$ bins in their respective SRs determined from the GPR estimates performed before this fit, and are used in a similar way to other backgrounds in this analysis. However, as the GPR predictions are posterior distributions over possible background estimates, the shapes and normalisations of this background are allowed to vary within the multi-dimensional Gaussian posterior distributions, which arise from these GPR estimates. Other systematic uncertainties are modelled as nuisance parameters and are profiled in the fit. The profile likelihood ratio test statistic~\cite{Cowan:2010js} is used to perform one-sided frequentist hypothesis tests under the asymptotic approximation~\cite{Cowan:2010js} to constrain the inclusive $H+c$ cross-section at the 95\% confidence level using the $\text{CL}_\text{s}$ formalism~\cite{Read:2002hq}, and to calculate the statistical significance of the inclusive $H+c$ signal. This $H+c$ cross-section is inclusive of decay phase space except as specified in the signal definition in section~\ref{sec:intro}, and does not include the branching fraction of the Higgs boson to photons. Expected results are calculated using an Asimov dataset~\cite{Cowan:2010js} that includes the expected signal.

\section{Results}

\begin{figure}[!htbp]
\centering
\subfloat[]{\includegraphics[width=0.49\textwidth]{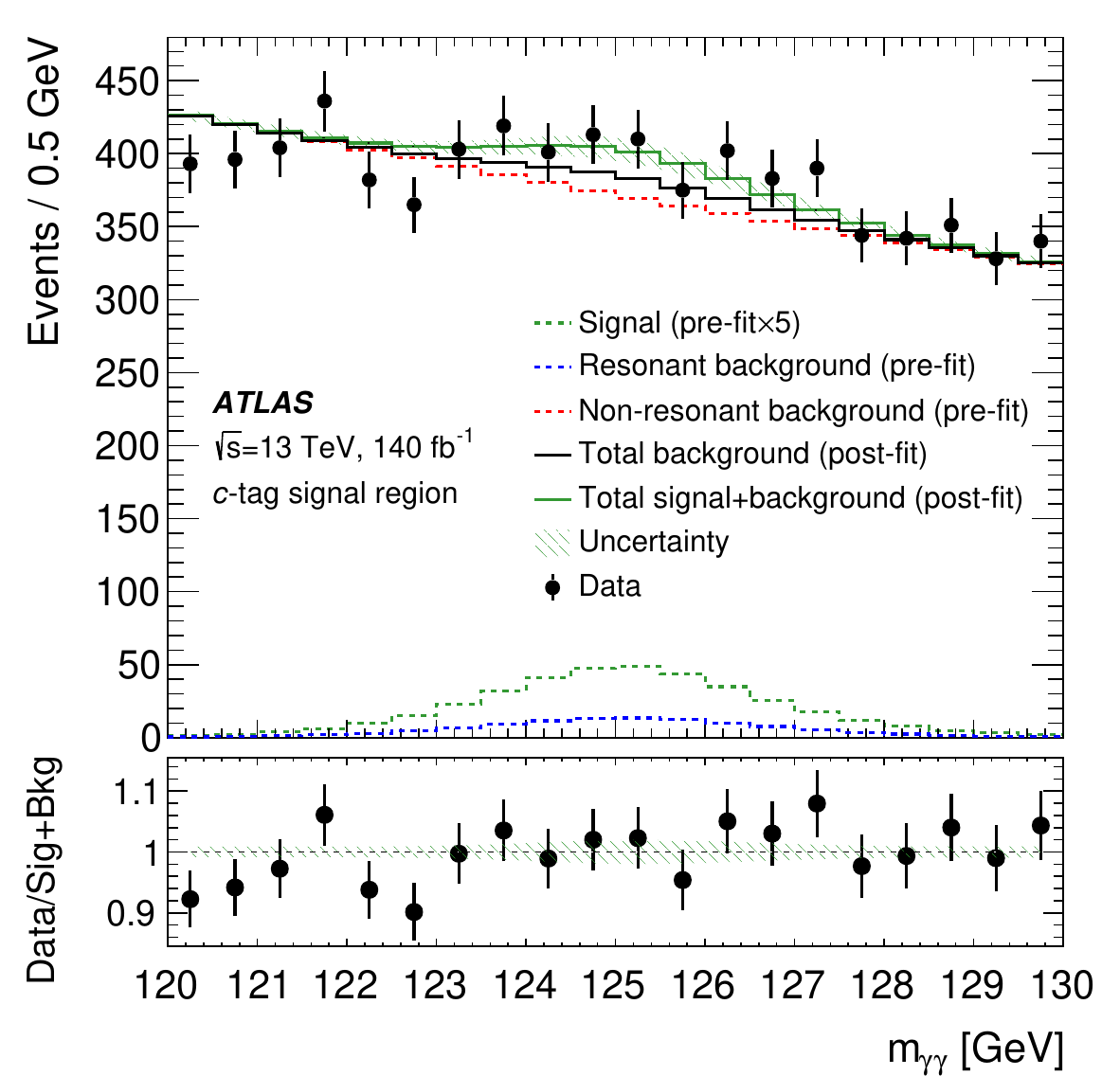}}
\subfloat[]{\includegraphics[width=0.49\textwidth]{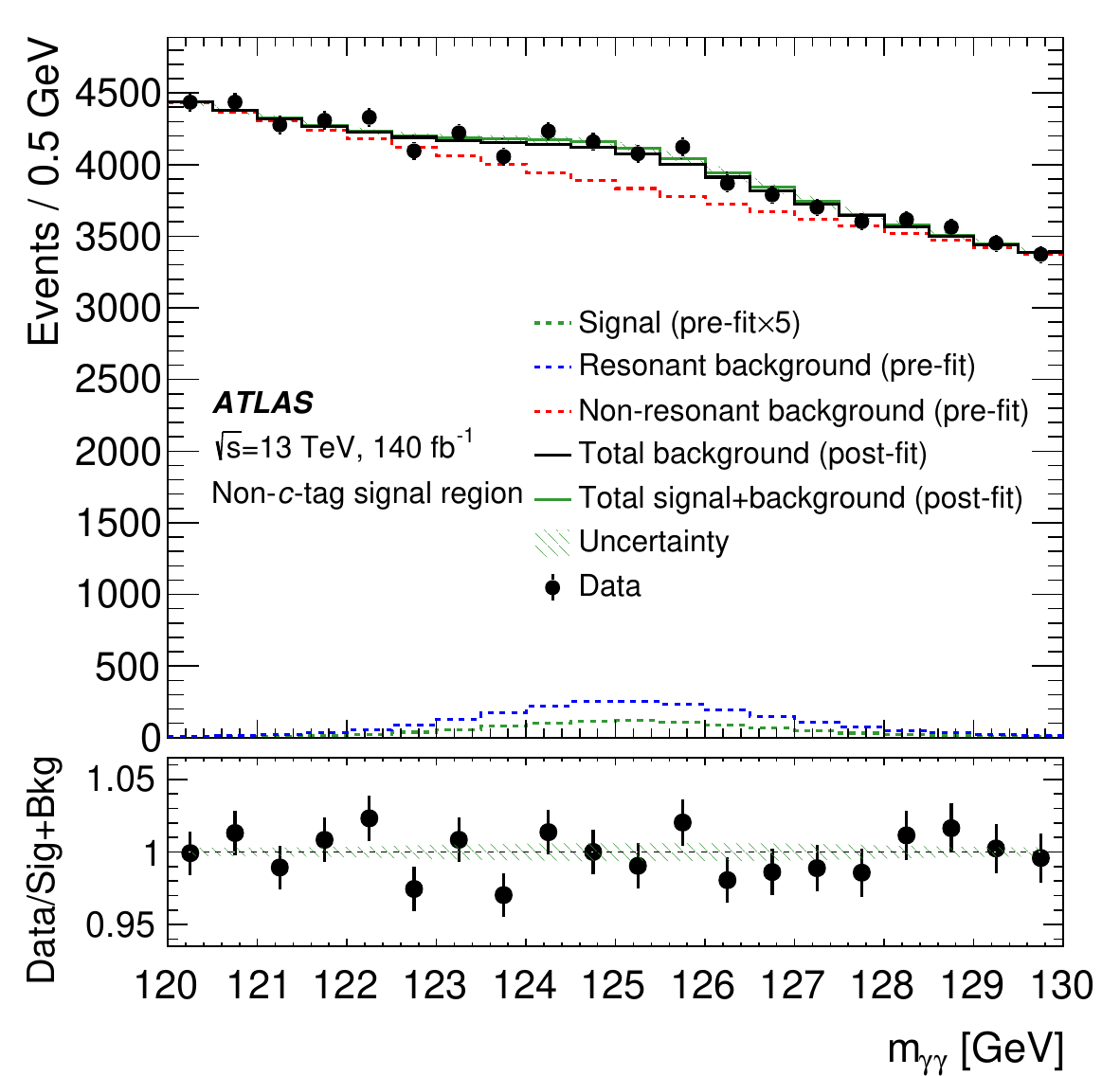}}
\caption{Pre- and post-fit $m_{\gamma\gamma}$ distributions in (a) the $c$-tag and (b) the non-$c$-tag signal regions. The pre-fit non-resonant background is estimated by using the Gaussian process regression-based interpolation, and the other pre-fit distributions are estimated by using simulation, as described in the main text. The hatched band represents the total post-fit uncertainty. The signal is scaled up by a factor of 5. The lower panel represents the ratio between the data and the total signal-plus-background prediction.}%
\label{fig:myy}
\end{figure}

The observed (expected) best-fit value of the inclusive $H+c$ signal cross-section is $5.3\pm 3.2$~pb ($2.9\pm 3.1$~pb), and the post-fit $m_{\gamma\gamma}$ distributions are shown in Figure~\ref{fig:myy}. The observed inclusive $H+c$ signal is compatible with the SM expectation, and the observed (expected) significance is $1.7\sigma$ ($1.0\sigma$). Observed (expected) upper limits are set on the inclusive $H+c$ cross-section at the 95\% confidence-level at 10.6~pb (8.8~pb) in the combined fit, and at 14.6~pb (9.6~pb) and 11.2~pb (14.1~pb) in the $c$-tag and non-$c$-tag SRs, respectively. The sensitivity of this analysis to the predicted normalisation of the total resonant Higgs boson background is tested by producing additional results in which this normalisation is left unconstrained, for which the best fit inclusive $H+c$ signal cross-section becomes $8.5\pm 4.0$~pb, the expected limit deteriorates by 18\% and the correlation between the signal and resonant background normalisations evaluated on the Asimov dataset is $-64\%$.

\FloatBarrier

\section{Conclusions}

A search for the production of a Higgs boson and one or more charm quarks using Higgs boson decays into photon pairs is presented. The search uses a data sample of $\sqrt{s}=13~\text{\TeV}$ proton-proton collision data collected between 2015 and 2018 with the ATLAS detector at the CERN LHC, amounting to an integrated luminosity of 140~fb$^{-1}$. This analysis provides is the first direct constraint on the inclusive $H+c$ cross-section, and achieves an expected signal uncertainty at approximately the same level as the expected signal.
This analysis also presents the first usage of Gaussian process regression directly for a background estimate in a particle physics data analysis, furthering pre-existing ideas of how to use this technique.
The measurement of this process is an important step towards probing the coupling of the Higgs boson to the charm quark using $H+c$ events, which would complement existing approaches to probe this coupling.
\FloatBarrier

\FloatBarrier

\section*{Acknowledgements}
%

%
%

%
%

We thank CERN for the very successful operation of the LHC and its injectors, as well as the support staff at
CERN and at our institutions worldwide without whom ATLAS could not be operated efficiently.

The crucial computing support from all WLCG partners is acknowledged gratefully, in particular from CERN, the ATLAS Tier-1 facilities at TRIUMF/SFU (Canada), NDGF (Denmark, Norway, Sweden), CC-IN2P3 (France), KIT/GridKA (Germany), INFN-CNAF (Italy), NL-T1 (Netherlands), PIC (Spain), RAL (UK) and BNL (USA), the Tier-2 facilities worldwide and large non-WLCG resource providers. Major contributors of computing resources are listed in Ref.~\cite{ATL-SOFT-PUB-2023-001}.

We gratefully acknowledge the support of ANPCyT, Argentina; YerPhI, Armenia; ARC, Australia; BMWFW and FWF, Austria; ANAS, Azerbaijan; CNPq and FAPESP, Brazil; NSERC, NRC and CFI, Canada; CERN; ANID, Chile; CAS, MOST and NSFC, China; Minciencias, Colombia; MEYS CR, Czech Republic; DNRF and DNSRC, Denmark; IN2P3-CNRS and CEA-DRF/IRFU, France; SRNSFG, Georgia; BMBF, HGF and MPG, Germany; GSRI, Greece; RGC and Hong Kong SAR, China; ISF and Benoziyo Center, Israel; INFN, Italy; MEXT and JSPS, Japan; CNRST, Morocco; NWO, Netherlands; RCN, Norway; MNiSW, Poland; FCT, Portugal; MNE/IFA, Romania; MSTDI, Serbia; MSSR, Slovakia; ARIS and MVZI, Slovenia; DSI/NRF, South Africa; MICIU/AEI, Spain; SRC and Wallenberg Foundation, Sweden; SERI, SNSF and Cantons of Bern and Geneva, Switzerland; NSTC, Taipei; TENMAK, T\"urkiye; STFC/UKRI, United Kingdom; DOE and NSF, United States of America.

Individual groups and members have received support from BCKDF, CANARIE, CRC and DRAC, Canada; CERN-CZ, FORTE and PRIMUS, Czech Republic; COST, ERC, ERDF, Horizon 2020, ICSC-NextGenerationEU and Marie Sk{\l}odowska-Curie Actions, European Union; Investissements d'Avenir Labex, Investissements d'Avenir Idex and ANR, France; DFG and AvH Foundation, Germany; Herakleitos, Thales and Aristeia programmes co-financed by EU-ESF and the Greek NSRF, Greece; BSF-NSF and MINERVA, Israel; NCN and NAWA, Poland; La Caixa Banking Foundation, CERCA Programme Generalitat de Catalunya and PROMETEO and GenT Programmes Generalitat Valenciana, Spain; G\"{o}ran Gustafssons Stiftelse, Sweden; The Royal Society and Leverhulme Trust, United Kingdom.

In addition, individual members wish to acknowledge support from Armenia: Yerevan Physics Institute (FAPERJ); CERN: European Organization for Nuclear Research (CERN PJAS); Chile: Agencia Nacional de Investigaci\'on y Desarrollo (FONDECYT 1230812, FONDECYT 1230987, FONDECYT 1240864); China: Chinese Ministry of Science and Technology (MOST-2023YFA1605700), National Natural Science Foundation of China (NSFC - 12175119, NSFC 12275265, NSFC-12075060); Czech Republic: Czech Science Foundation (GACR - 24-11373S), Ministry of Education Youth and Sports (FORTE CZ.02.01.01/00/22\_008/0004632), PRIMUS Research Programme (PRIMUS/21/SCI/017); EU: H2020 European Research Council (ERC - 101002463); European Union: European Research Council (ERC - 948254, ERC 101089007), Horizon 2020 Framework Programme (MUCCA - CHIST-ERA-19-XAI-00), European Union, Future Artificial Intelligence Research (FAIR-NextGenerationEU PE00000013), Italian Center for High Performance Computing, Big Data and Quantum Computing (ICSC, NextGenerationEU); France: Agence Nationale de la Recherche (ANR-20-CE31-0013, ANR-21-CE31-0013, ANR-21-CE31-0022, ANR-22-EDIR-0002), Investissements d'Avenir Labex (ANR-11-LABX-0012); Germany: Baden-Württemberg Stiftung (BW Stiftung-Postdoc Eliteprogramme), Deutsche Forschungsgemeinschaft (DFG - 469666862, DFG - CR 312/5-2); Italy: Istituto Nazionale di Fisica Nucleare (ICSC, NextGenerationEU), Ministero dell'Università e della Ricerca (PRIN - 20223N7F8K - PNRR M4.C2.1.1); Japan: Japan Society for the Promotion of Science (JSPS KAKENHI JP22H01227, JSPS KAKENHI JP22H04944, JSPS KAKENHI JP22KK0227, JSPS KAKENHI JP23KK0245); Netherlands: Netherlands Organisation for Scientific Research (NWO Veni 2020 - VI.Veni.202.179); Norway: Research Council of Norway (RCN-314472); Poland: Ministry of Science and Higher Education (IDUB AGH, POB8, D4 no 9722), Polish National Agency for Academic Exchange (PPN/PPO/2020/1/00002/U/00001), Polish National Science Centre (NCN 2021/42/E/ST2/00350, NCN OPUS nr 2022/47/B/ST2/03059, NCN UMO-2019/34/E/ST2/00393, UMO-2020/37/B/ST2/01043, UMO-2021/40/C/ST2/00187, UMO-2022/47/O/ST2/00148, UMO-2023/49/B/ST2/04085, UMO-2023/51/B/ST2/00920); Slovenia: Slovenian Research Agency (ARIS grant J1-3010); Spain: Generalitat Valenciana (Artemisa, FEDER, IDIFEDER/2018/048), Ministry of Science and Innovation (MCIN \& NextGenEU PCI2022-135018-2, MICIN \& FEDER PID2021-125273NB, RYC2019-028510-I, RYC2020-030254-I, RYC2021-031273-I, RYC2022-038164-I), PROMETEO and GenT Programmes Generalitat Valenciana (CIDEGENT/2019/027); Sweden: Swedish Research Council (Swedish Research Council 2023-04654, VR 2018-00482, VR 2022-03845, VR 2022-04683, VR 2023-03403, VR grant 2021-03651), Knut and Alice Wallenberg Foundation (KAW 2018.0157, KAW 2018.0458, KAW 2019.0447, KAW 2022.0358); Switzerland: Swiss National Science Foundation (SNSF - PCEFP2\_194658); United Kingdom: Leverhulme Trust (Leverhulme Trust RPG-2020-004), Royal Society (NIF-R1-231091); United States of America: U.S. Department of Energy (ECA DE-AC02-76SF00515), Neubauer Family Foundation.

%
%


%
%
%
%
%
%
%

%
%
%
%
%
%

%
%

%
%
%
\printbibliography
\clearpage
 
\begin{flushleft}
\hypersetup{urlcolor=black}
{\Large The ATLAS Collaboration}

\bigskip

\AtlasOrcid[0000-0002-6665-4934]{G.~Aad}$^\textrm{\scriptsize 104}$,
\AtlasOrcid[0000-0001-7616-1554]{E.~Aakvaag}$^\textrm{\scriptsize 17}$,
\AtlasOrcid[0000-0002-5888-2734]{B.~Abbott}$^\textrm{\scriptsize 123}$,
\AtlasOrcid[0000-0002-0287-5869]{S.~Abdelhameed}$^\textrm{\scriptsize 119a}$,
\AtlasOrcid[0000-0002-1002-1652]{K.~Abeling}$^\textrm{\scriptsize 56}$,
\AtlasOrcid[0000-0001-5763-2760]{N.J.~Abicht}$^\textrm{\scriptsize 50}$,
\AtlasOrcid[0000-0002-8496-9294]{S.H.~Abidi}$^\textrm{\scriptsize 30}$,
\AtlasOrcid[0009-0003-6578-220X]{M.~Aboelela}$^\textrm{\scriptsize 45}$,
\AtlasOrcid[0000-0002-9987-2292]{A.~Aboulhorma}$^\textrm{\scriptsize 36e}$,
\AtlasOrcid[0000-0001-5329-6640]{H.~Abramowicz}$^\textrm{\scriptsize 154}$
% ...
\bigskip
\\

% ...
$^{17}$Department for Physics and Technology, University of Bergen, Bergen; Norway.\\
% ...
$^{30}$Physics Department, Brookhaven National Laboratory, Upton NY; United States of America.\\
% ...
$^{36}$$^{(a)}$Facult\'e des Sciences Ain Chock, Universit\'e Hassan II de Casablanca;$^{(b)}$Facult\'{e} des Sciences, Universit\'{e} Ibn-Tofail, K\'{e}nitra;$^{(c)}$Facult\'e des Sciences Semlalia, Universit\'e Cadi Ayyad, LPHEA-Marrakech;$^{(d)}$LPMR, Facult\'e des Sciences, Universit\'e Mohamed Premier, Oujda;$^{(e)}$Facult\'e des sciences, Universit\'e Mohammed V, Rabat;$^{(f)}$Institute of Applied Physics, Mohammed VI Polytechnic University, Ben Guerir; Morocco.\\
% ...
$^{45}$Physics Department, Southern Methodist University, Dallas TX; United States of America.\\
% ...
$^{50}$Fakult\"{a}t Physik , Technische Universit{\"a}t Dortmund, Dortmund; Germany.\\
% ...
$^{56}$II. Physikalisches Institut, Georg-August-Universit\"{a}t G\"ottingen, G\"ottingen; Germany.\\
% ...
$^{104}$CPPM, Aix-Marseille Universit\'e, CNRS/IN2P3, Marseille; France.\\
% ...
$^{119}$$^{(a)}$New York University Abu Dhabi, Abu Dhabi;$^{(b)}$United Arab Emirates University, Al Ain; United Arab Emirates.\\
% ...
$^{123}$Homer L. Dodge Department of Physics and Astronomy, University of Oklahoma, Norman OK; United States of America.\\
% ...
$^{154}$Raymond and Beverly Sackler School of Physics and Astronomy, Tel Aviv University, Tel Aviv; Israel.\\
% ...
$^{a}$ Also Affiliated with an institute covered by a cooperation agreement with CERN.\\
$^{b}$ Also at An-Najah National University, Nablus; Palestine.\\
$^{c}$ Also at Borough of Manhattan Community College, City University of New York, New York NY; United States of America.\\
$^{d}$ Also at Center for Interdisciplinary Research and Innovation (CIRI-AUTH), Thessaloniki; Greece.\\
$^{e}$ Also at CERN, Geneva; Switzerland.\\
$^{f}$ Also at CMD-AC UNEC Research Center, Azerbaijan State University of Economics (UNEC); Azerbaijan.\\
% ...

\end{flushleft}


%

%
%

%

%
%
%
%

\end{document}